\documentclass[twocolumn]{aastex631}

\usepackage{amsmath}
\usepackage{graphicx}
\usepackage{multirow}

\shorttitle{Anisotropic Diffusion Model for Pulsar Halos}
\shortauthors{Yan et al.}

\begin{document}

\title{Prospect of detecting TeV halos with LHAASO:  in the framework of the anisotropic diffusion model}

\author[0000-0001-7542-5861]{Kai Yan}
\affiliation{School of Astronomy and Space Science, Nanjing University, Nanjing 210023, China}

\author{Ruo-Yu Liu}
\affiliation{School of Astronomy and Space Science, Nanjing University, Nanjing 210023, China}
\affiliation{Key laboratory of Modern Astronomy and Astrophysics (Nanjing University), Ministry of Education, Nanjing 210023, China}

\author{S.Z. Chen}
\affiliation{Key Laboratory of Particle Astrophyics, Institute of High Energy Physics, Chinese Academy of Sciences,  Beijing 100049, China}
\affiliation{TIANFU Cosmic Ray Research Center, Chengdu, Sichuan, China}

\author{Xiang-Yu Wang}
\affiliation{School of Astronomy and Space Science, Nanjing University, Nanjing 210023, China}
\affiliation{Key laboratory of Modern Astronomy and Astrophysics (Nanjing University), Ministry of Education, Nanjing 210023, China}

\correspondingauthor{Ruo-Yu Liu}
\email{ryliu@nju.edu.cn}

\begin{abstract}
The particle diffusion coefficients of three TeV pulsar halos observed so far are inferred to be significantly smaller than the typical value of the interstellar medium (ISM). The anisotropic diffusion model ascribes the slow diffusion to the cross-field diffusion assuming sub-Alfv{\'e}nic turbulence in the ISM around the pulsar if the viewing angle between the observer's line-of-sight (LOS) to the pulsar and the local mean field direction is small. In general, the TeV halo's morphology under this model highly depends on the viewing angle, and an elongated, asymmetric morphology is predicted if the LOS is not approximately aligned with the local mean field direction. While the specific requirement of a small viewing angle is supposedly established only for a small fraction of TeV halos, TeV halos with apparent asymmetric morphology has not been detected. In this paper we will study the expectation of TeV halos measured by the TeV-PeV gamma-ray detector LHAASO in the framework of anisotropic diffusion model, with a particular focus on the influence of the viewing angle on the detectability. We show that a TeV halo is more detectable with a smaller viewing angle and this selection effect may explain why the morphologies of all three detected TeV halos so far are consistent being spherical. We also demonstrate that LHAASO is capable of detecting asymmetric TeV halos after several-year operation with reasonable source parameters. This can serve as a critical test of the anisotropic diffusion model.
\end{abstract}

%\keywords{TeV halos; anisotropic diffusion; LHAASO observation}

\section{Introduction}

The Large High-Altitude Air Shower Observatory (LHAASO) has recently discovered an extended gamma-ray source around the middle-aged pulsar PSR J0622+3749 above 25\,TeV \citep{aharonian2021extended}. This source, termed as LHAASO J0621+3755, is quite similar to the so-called TeV pulsar halos previously identified by the High Altitude Water Cherenkov (HAWC) telescope around the Geminga pulsar and the Monogem pulsar \citep{abeysekara2017extended}. TeV halos are the relativistic electron-rich region around middle-aged pulsars, where pulsars no longer dominate the dynamics \citep{2020A&A...636A.113G}. A TeV halo results from electrons and positrons, which have escaped from the pulsar wind nebulae into the ambient interstellar medium (ISM), up-scattering the cosmic microwave backgrond (CMB) and the interstellar infrared radiation field \citep{2017PhRvD..96j3016L, 2019ApJ...879...91J, 2019PhRvD.100d3016S, 2020PhRvD.101j3035D}. For photons above 10\,TeV,  the infrared radiation field becomes less important than CMB due to the Klein-Nishina effect and the relation between the energy of emitting electron (hereafter we do not distinguish positrons from electrons for simplicity)  $E_e$ and the energy of the up-scattered photon $E_\gamma$ can be given by $E_e\approx 100\, (E_\gamma/20{\rm TeV})^{1/2}\,$TeV.

All the three measured TeV halos are spatially extended over at least $20-30$\,pc. Based on their multi-TeV intensity profiles, the inferred spatial distribution of electrons at the small radius declines more steeply than $r^{-1}$, implying that escaping electrons cool before diffusing to a large distance. Employing a simple isotropic diffusion model, a small diffusion coefficient of $D\lesssim 10^{28}\rm cm^{2}s^{-1}$ at electron energy $E_{\rm e} \sim 100 \rm TeV$ \citep{abeysekara2017extended,aharonian2021extended} is required to fit the intensity profile provided the typical magnetic field strength of the ISM. The obtained diffusion coefficient is significantly smaller than that inferred from the measurements on the local secondary-to-primary cosmic-ray (CR) ratios by about two orders of magnitude \citep[e.g.][]{2010ApJ...722L..58S}. The magnetic field in pulsar halos may be weaker than the typical ISM value according to the X-ray observation \citep{2019ApJ...875..149L}, leading to the requirement of an even smaller diffusion coefficient.
%($\sim 10^{30}-10^{31}cm^{2}s^{-1}$) \citep{2011ApJ...729..106T,2021ChPhL..38c9801F}. 

Such a slow diffusion zone needs to extend at least a few tens parsecs around the pulsar to account for the observation, but its origin is still unclear. \citet{2018MNRAS.479.4526L} found that the injection scale of the external turbulence need be $\lesssim 1$\,pc to produce the slow diffusion coefficient, which, however, is much smaller than the typical injection scale of the interstellar turbulence. There are also suggestions that electrons in the halo may amplify the turbulent magnetic field therein via the streaming instability and restrain their own transport \citep{2018Evoli,2021arXiv211101143M}, but the amount of the injected electrons may not be sufficient to reduce the diffusion coefficient down to the required level as the pulsars are already in their middle ages \citep{2019MNRAS.488.4074F}. This becomes even more problematic at higher energies, in particular at $E_e>100\,$TeV, due to the rapid cooling of these extremely energetic electrons. The related supernova remnants might be a possible origin of the slow diffusion, either by the shocks \citep{2019MNRAS.488.4074F} or by the accelerated protons \citep{2021arXiv211101143M}, although a dedicated demonstration is yet to be performed. 

Alternatively, \citet{PhysRevLett.123.221103} proposed an anisotropic diffusion model assuming the turbulence of the ambient ISM to be sub-Alfv{\'e}nic with the Alfv{\'e}nic Mach number $M_A$ less than unity. In this model, the electron diffusion perpendicular to the mean magnetic field can be much slower than that along the mean magnetic field. The spatial distribution of the electron density is not spherically symmetric but instead in a cylindrical symmetry with respect to the mean magnetic field direction. Given such a spatial distribution, the viewing angle $\phi$, which is defined as the angle between the observer's LOS towards the pulsar and the mean magnetic field direction, is relevant with the apparent morphology of the halo. A small viewing angle (i.e., $\lesssim 5^\circ$) is needed to reproduce the roughly spherical morphology of the halo and to make the electron diffusion look slow in the plane of sky \citep{PhysRevLett.123.221103}. 

While the anisotropic diffusion model can naturally explain the slow diffusion and the nondetection of the X-ray emission from Geminga's pulsar halo, it requires a specific viewing angle of the halo. The morphology of two other detected TeV halos, the Monogem pulsar halo and LHAASO~J0621+3755, are also consistent with being spherically symmetric, although the Monogem halo appears somewhat elongated according to the updated observation of HAWC \citep{Zhou19}. If we assume TeV halos are common phenomena around middle-aged pulsars, TeV halos should not always appear spherical under the anisotropic diffusion model. It is not clear how are the directions of the mean magnetic field of the ISM around those pulsars distributed, but if we simply assume that they can be oriented to any direction with an equal probability, the probability of finding a pulsar halo with a viewing angle smaller than a critical value $\phi_c$ can be estimated by $\phi_c^2/2=0.004(\phi_c/5^\circ)^2$. As a result, if we can observe every TeV halos, most of them are supposed to show asymmetric or elongated morphology in the anisotropic diffusion model. On the other hand, however, pulsar halos with smaller viewing angles should be easier to be detected than those with larger viewing angles provided other conditions are the same. This is because the former would appear more compact and bright in the sky, resulting in a selection effect inclined to see spherically symmetric halos or halos with small viewing angles.

The morphology of TeV pulsar halos is a key observable to distinguish between the anisotropic diffusion model and the isotropic diffusion models. Therefore, it is important to look into the selection effect in a more quantitative way, especially the influence from the viewing angle. In this paper, we will study the detection capacity of LHAASO's kilometer square array (KM2A) for pulsar halo under the framework of anisotropic diffusion model. LHAASO has now operated over 10\,months with its full array. It is an ideal detector for extended sources above 10\,TeV given its large field of view and good sensitivity. Another reason of exploring the expected measurement of LHAASO-KM2A is that we may safely ignore the influence of the pulsar's proper motion and the uncertainty of the spindown history of middle-aged pulsars because electrons emitting at such high energies cool rapidly in typical interstellar magnetic field and the CMB radiation field \citep{2021ApJ...922..130Z}. We will show that LHAASO is hopeful to distinguish the two models for the pulsar halo after a few year's operation.

The rest of the paper is organized as follows: in Section 2, we will introduce the method to obtain the two-dimensional (2D) intensity profile of pulsar halos in the anisotropic diffusion model and evaluate the detectability of LHAASO on these sources; in Section 3, we present and discuss the main results of the paper; we further discuss some uncertainties of model parameters in Section 4 and give our conclusion in Section 5.

\begin{figure*}[htbp]
\hspace{-1.9cm} \includegraphics[width=1.2\textwidth]{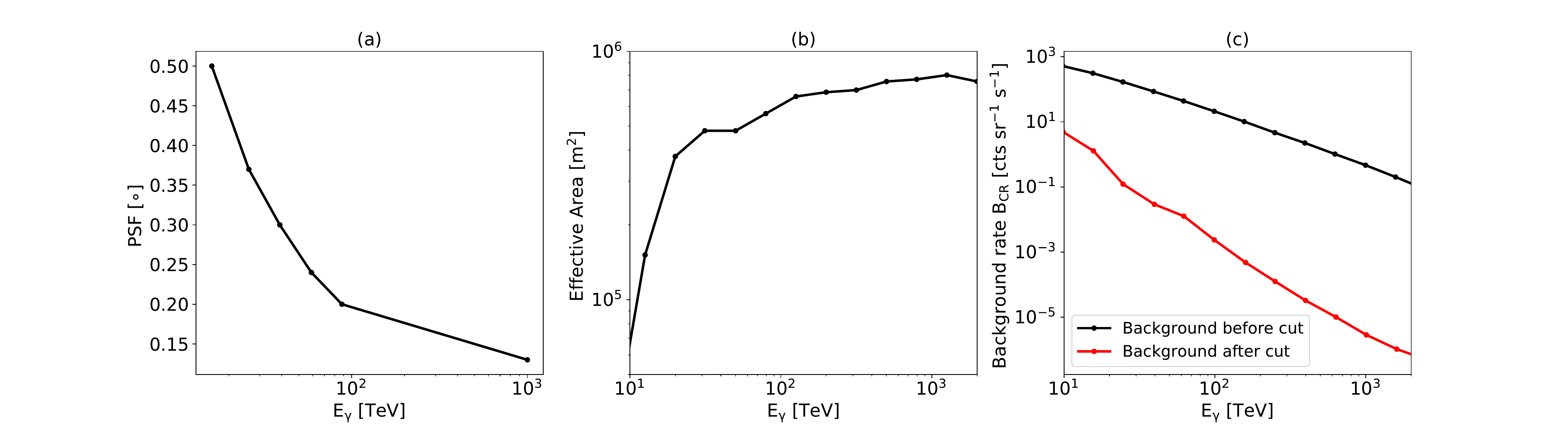}
\caption{(a) the size of PSF of LHAASO-KM2A as a function of energy. (b) the effective area of LHAASO-KM2A at different energies for an average zenith angle of $30^{\circ}$. (c) the background counts rate before (black) and after (red) muon cut above different energies. The red curve represents $B_{\rm CR}$.}
\label{fig:performance}
\end{figure*}

\section{Methods} \label{sec:style}
\subsection{Spatial Distribution of Electrons and the Radiation}
The transport of escaping electrons in a TeV halo is dominated by diffusion. We write the transport equation in the cylindrical coordinate and take the direction of the mean magnetic field as the $z$-axis, following \citet{PhysRevLett.123.221103}, i.e.,
\begin{equation}
\begin{aligned}
\label{E1}
\frac{\partial N}{\partial t}=&\frac{1}{r} \frac{\partial}{\partial r}\left(r D_{r r} \frac{\partial N}{\partial r}\right)+D_{z z} \frac{\partial^{2} N}{\partial z^{2}} \\&
-\frac{\partial}{\partial E_{\rm e}}\left(\dot{E}_{e} N\right)+Q\left(E_{e}\right) s(t) \delta(r) \delta(z)
\end{aligned}
\end{equation}
where $N\equiv N(E_e,r,t)$ is the particle number density, $D_{zz}$ and $D_{rr}$ are the diffusion coefficient parallel and perpendicular to the mean magnetic field direction respectively. $D_{zz}=D_{0}(E_{e}/\mathrm{1GeV})^{q}$ with $q=1/3$ following the Kolmogorov's theory and $D_{0}$ is taken to be $\mathrm{10^{28}cm^{2}s^{-1}}$ as the standard value in ISM. $D_{rr}=D_{zz}M_{\rm A}^{4}$ \citep{2008ApJ...673..942Y}. $\dot{E}_{e}$ is the cooling rate of electrons through the inverse Compton (IC) radiation in the CMB and the interstellar infrared radiation field with a temperature of $30\,$K and an energy density of $0.3\rm \,eV/cm^3$, and through the synchrontron radiation in the interstellar magnetic field of the strength $B=5\mu$G. The last term in the equation represents the injection term from the pulsar wind nebula which is approximated as a point source. $Q(E_{e})$ is the injection spectrum of electrons, denoted as $Q(E_{e})=N_{0}E_{e}^{-p}e^{-E_{e}/E_{\rm max}}$, where $p$ is the spectral index and and $E_{\rm max}$ is the cutoff energy in the spectrum. The normalization constant $N_{0}$ can be determined by
\begin{equation}
\label{E2}
\int_{E_{0}}^{\infty} E_{e} Q(E_{e})d E_{e} = \eta_{e} L_s
\end{equation}
where $\eta_{e}$ is an input parameter representing the ratio of pulsar spin-down energy that goes into pairs, and $L_{\rm s}$ is the pulsar total spin-down power. We set a minimum energy $E_{0}=0.1\,$TeV in the injection spectrum noting that this value is not very important since we mainly focus on $>10\,$TeV electrons which radiate in the observational energy range of LHAASO-KM2A. $s(t)$ is the temporal behavior of injection rate and denoted as $s(t)=(1+t/\tau)^{-2}$ with $\tau$ being the spin-down timescale of Geminga. $\delta(r)$ and $\delta(z)$ are Dirac functions denoting the injection location. 

After obtaining $N(E_e,r,t)$, we can calculate the IC emissivity of electrons $q_{\rm IC}(E,z,r)$ following the semi-analytical method given by \citet{2014ApJ...783..100K}. The 2D gamma-ray intensity profile, $I_\gamma (E,\theta, \xi)$, is the projection of the emission onto the plane-of-the-sky and can be obtained by integrating the emissivity over the observer's LOS towards each direction, as detailed in \citet{PhysRevLett.123.221103}.

% The intensity of IC radiation is calculated by
% \begin{equation}
% \begin{aligned}
% \label{E3}
% I_{\gamma }(E,\theta,\xi)=\frac{1}{4 \pi} \int_{l_{\min }}^{l_{\max }} \mathcal{F}_{\mathrm{IC}}\left\{N(E, z, r), n_{\mathrm{ph}}\right\} dl
% \end{aligned}
% \end{equation}
% where $n_{ph}$ is the differential density of the radiation backgrounds. $\mathcal{F}_{\mathrm{IC}}\left\{N(E, z, r), n_{\mathrm{ph}}\right\}$ is computed following \citet{2014ApJ...783..100K}
% \begin{equation}
% \begin{aligned}
% \label{E4}
% \mathcal{F}_{\mathrm{IC}} = &
% \sum_{i} \int \frac{2 r_{o}^{2} m_{\mathrm{e}}^{3} c^{4} \kappa_{i} T_{i}^{2}}{\pi \hbar^{3} E_{e}^{2}} \\ & \left[\frac{z^{2}}{2(1-z)} G_{3}\left(x_{0}\right)+G_{4}\left(x_{0}\right)\right]N(E_{e},z,r)dE_{e}
% \end{aligned}
% \end{equation}
% The sum in this equation is applied to different components of the radiation backgrounds, including the cosmic microwave background (CMB), infrared emission (IR), starlight, and ultraviolet emission (UV). $T_{\rm i}$ graybody emission temperature of the radiation backgrounds: $T_{\rm CMB}=$ 2.73K, $T_{\rm IR}=$ 20K, $T_{\rm star}=$5000K and $T_{\rm CMB}=$ 20,000K. $\kappa_{\rm i}$ is the corresponding dilution coefficient. $z=E/E_{e}$, and $x_{0}=z/4E_{\rm e}T_{\rm ph}(1-z)$. $G_{3,4}$ takes the same form as \citet{2014ApJ...783..100K}. 

\subsection{Estimation of LHAASO's detection}
To evaluate the LHAASO's performance on detection of a simulated pulsar halo, we need to estimate the signal-to-noise ratio (SNR) or the statistical significance of the halo's emission. Firstly, we need to convolve the theoretical 2D intensity profile with the point-spread-function (PSF) of LHAASO as shown in Fig.~\ref{fig:performance}a, which may modify the morphology of the halo observed by the instrument. Following \citet{2021ApJ...922..130Z}, a 2D Gaussian function is used to represent the PSF of LHAASO, so that the PSF-convolved 2D intensity profile can be given by
\begin{equation}
\begin{aligned}
\label{E5}
I_{\gamma,\rm PSF} =  \iint \frac{1}{2\pi \sigma ^{2}}\exp\left(-\frac{l^{'2}}{2\sigma^{2}}\right) I_{\gamma }^{'}(E_{\gamma},\theta^{'},\xi^{'})\sin\theta^{'}d\theta^{'}d\xi^{'} \end{aligned}
\end{equation}
where the angular distance $l^{'}=\arccos[\cos\theta \cos\theta^{'}+\sin\theta \sin\theta^{'}\cos(\xi -\xi^{'})]$ and $\sigma$ is the size of PSF as listed in Fig.~\ref{fig:performance}a. The expected photon counts rate above $E_\gamma$ per solid angle towards the direction ($\theta$, $\xi$) from the pulsar's position can be given by 
\begin{equation}
C_\gamma(>E_\gamma, \theta, \xi)=\int_{E_\gamma}^\infty A_{\rm eff}(E_\gamma)I_{\gamma,\rm PSF}/E_\gamma^2dE_\gamma,
\end{equation}
where $A_{\rm eff}$ is the effective area of LHAASO-KM2A as shown in Fig.~\ref{fig:performance}b. We follow the effective area of the half-KM2A array shown in \citet{2021ChPhC..45b5002A} and multiply it by a factor of 2 for $A_{\rm eff}$, assuming an average zenith angle of $30^\circ$. 

\begin{figure}[htbp]
\centering
\includegraphics[width=0.5\textwidth]{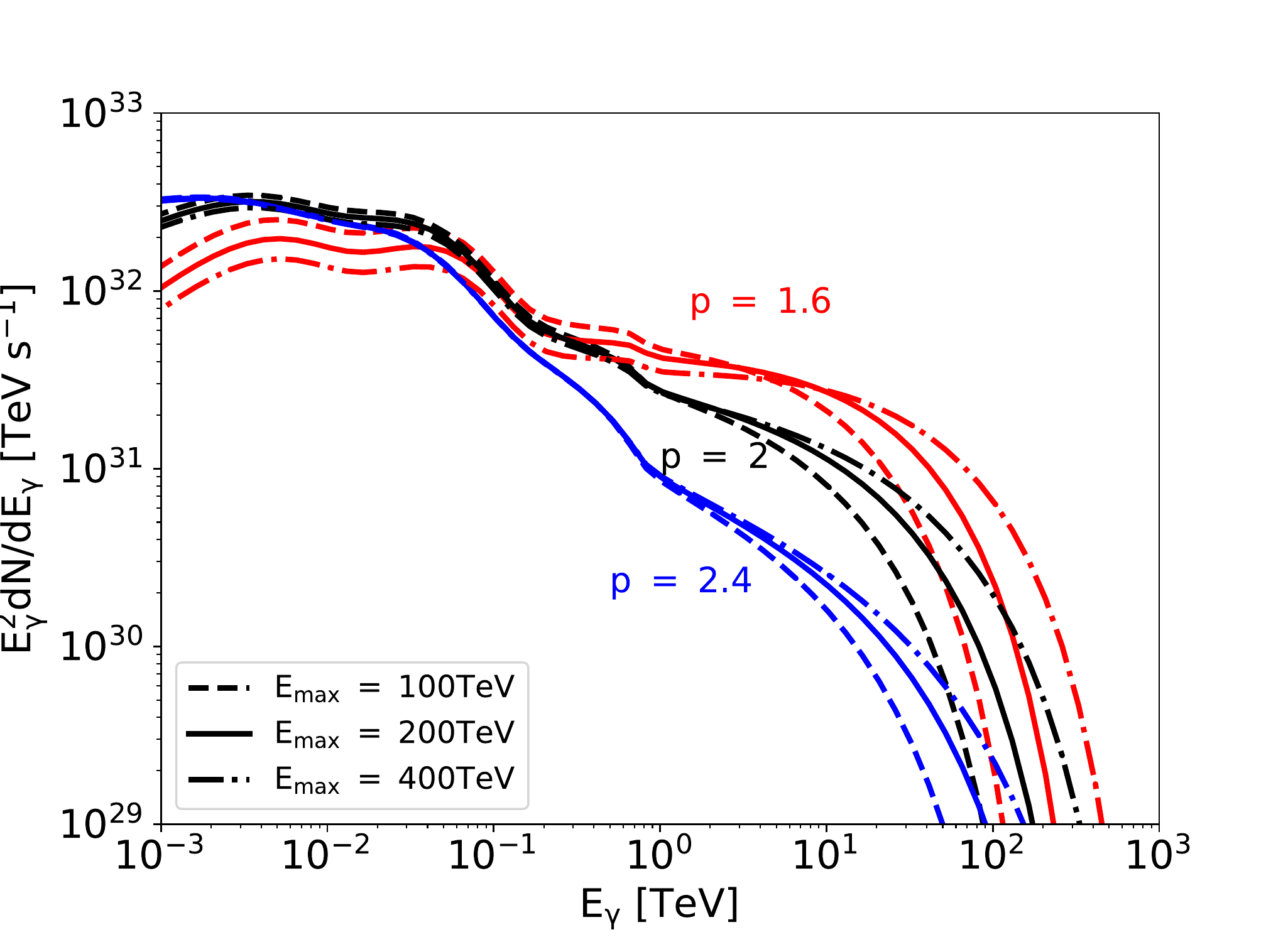}
\caption{Spectral energy distribution (SED) of the simulated halo at energy between $10^{-3}$\,TeV and $10^{3}$\,TeV. Different colors of curves represent different spectral index $p$. The red, black and blue curves represent $p$ = 1.6, 2, 2.4, respectively. Dashed, solid and dash-dotted curves represent $E_{\rm max}$ = 100, 200, 400 TeV, respectively. }
\label{fig:spectrum}
\end{figure}

\begin{figure*}[htbp]
\centering
    \includegraphics[width=0.45\textwidth]{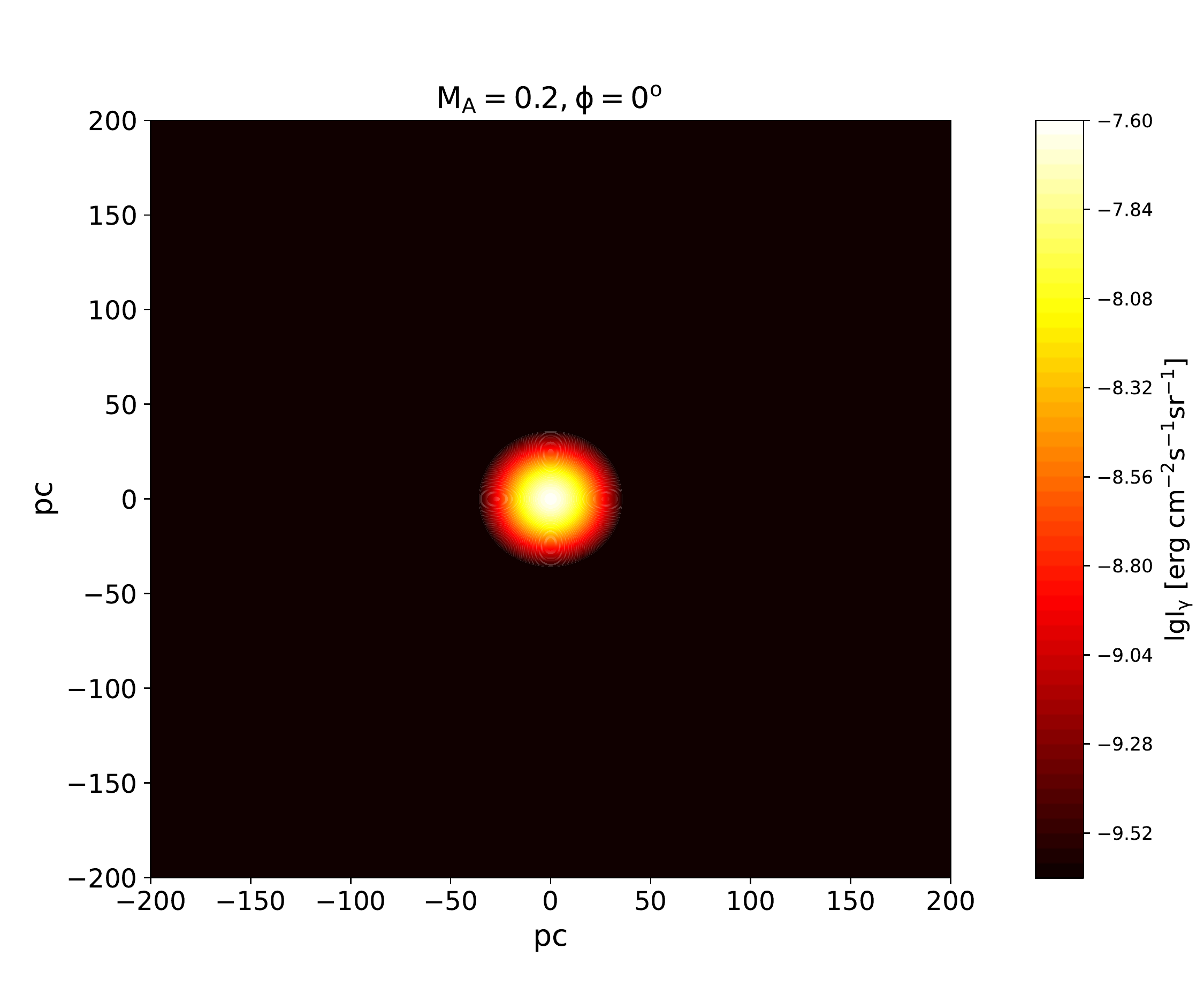}
    \includegraphics[width=0.45\textwidth]{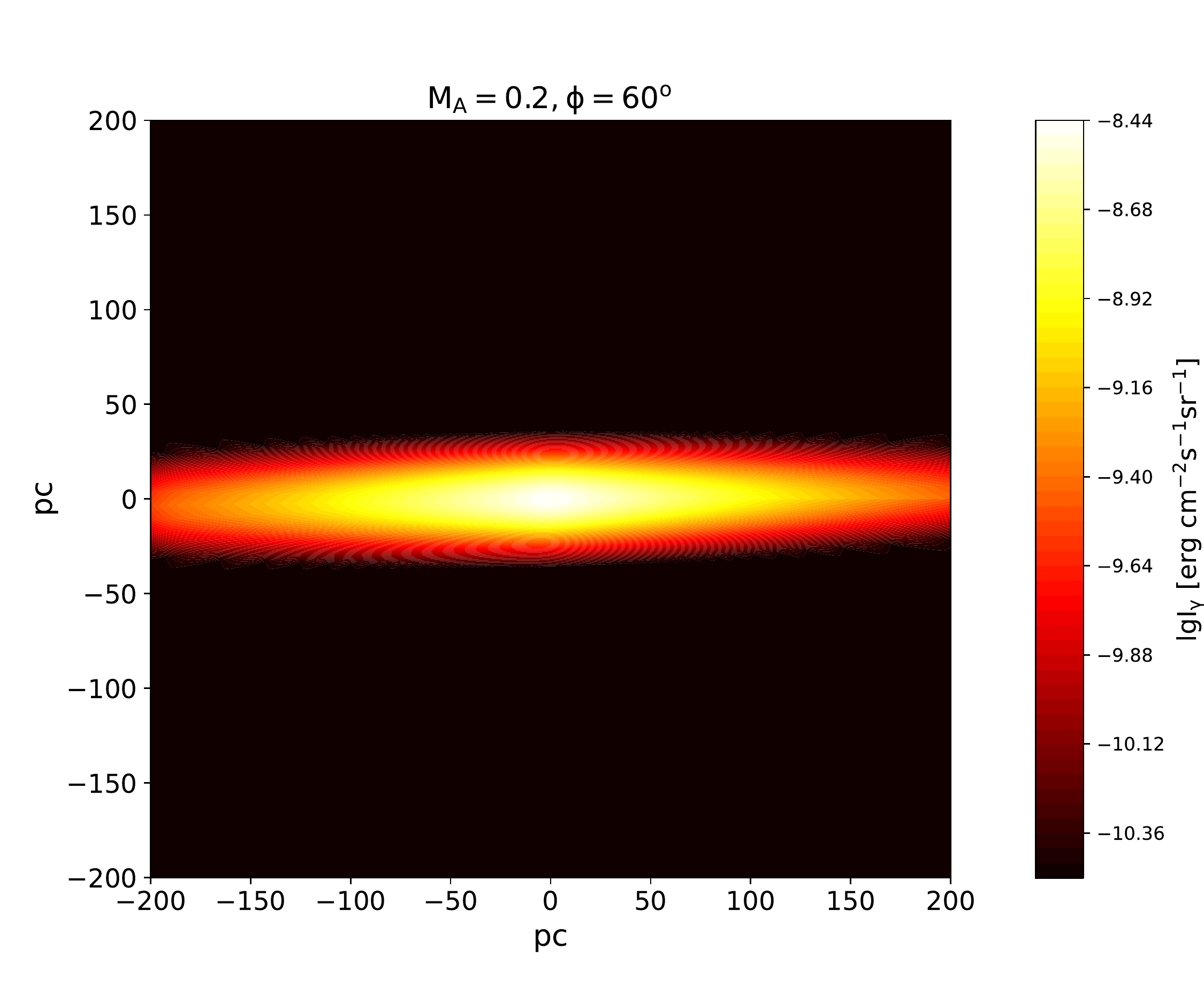}\\
    \includegraphics[width=0.45\textwidth]{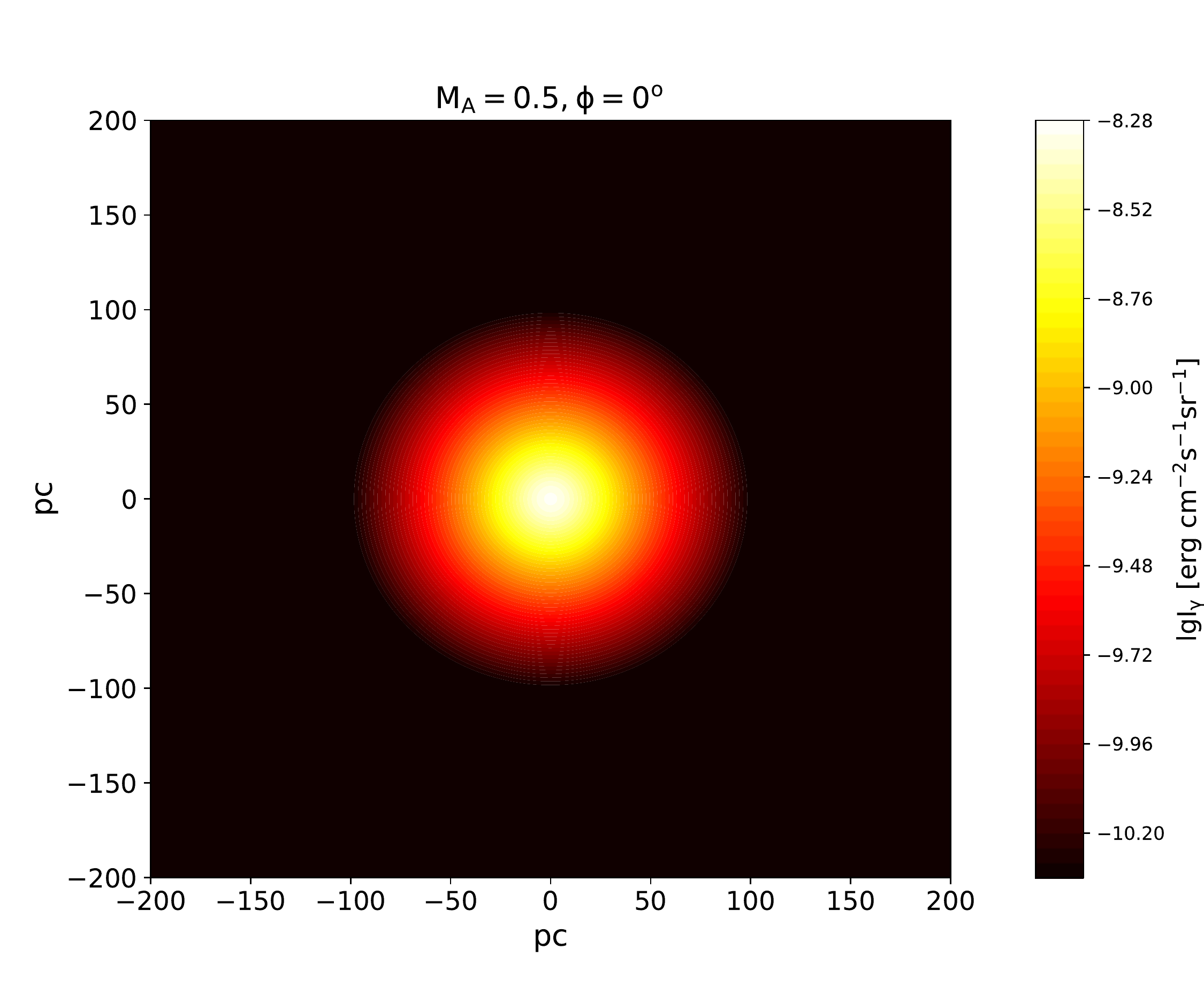}
    \includegraphics[width=0.45\textwidth]{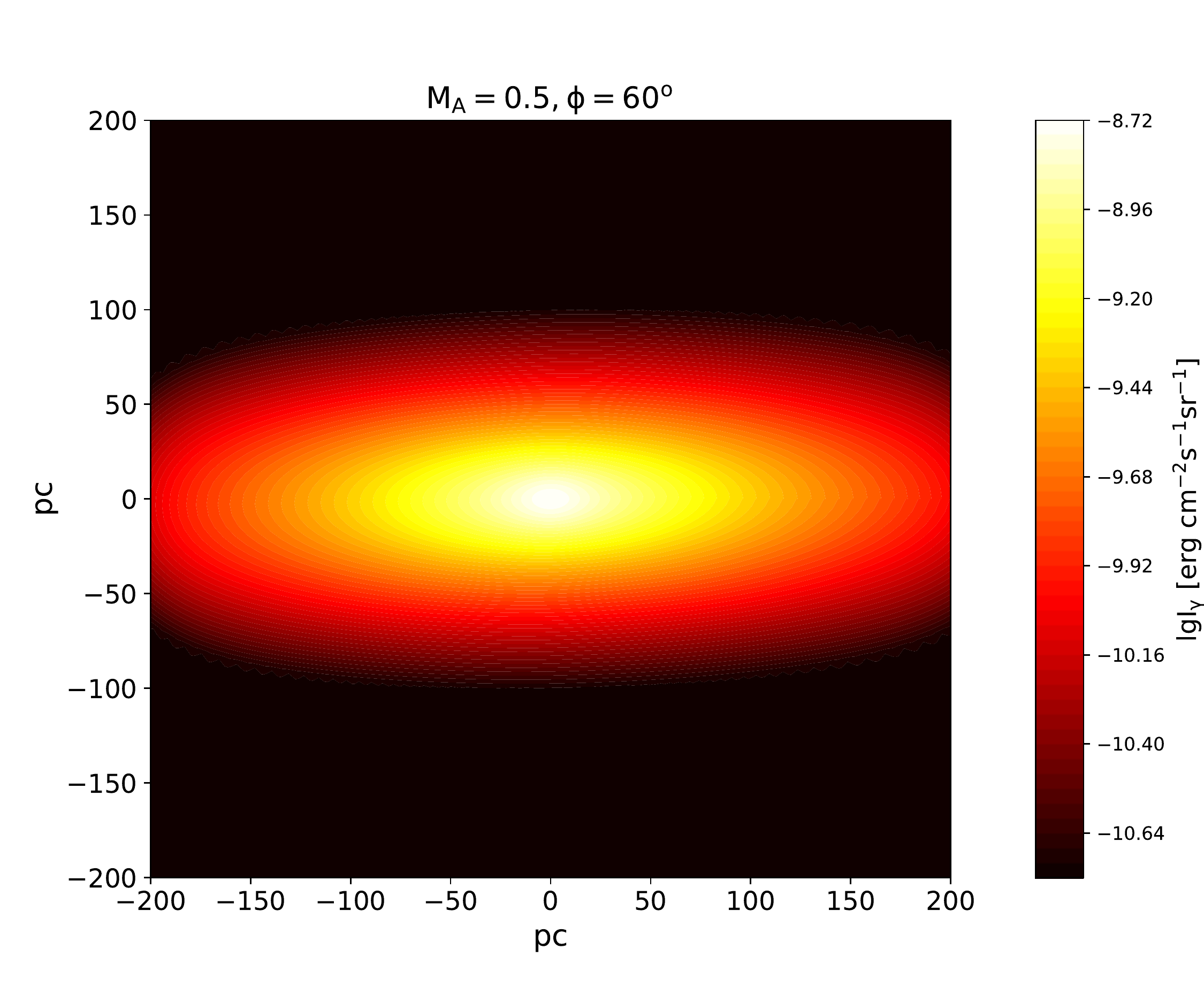}
\caption{Simulated 2D intensity profile in $25-1000$\,TeV  with $\mathrm{Alfv\acute{e}nic}$ Mach number $M_{\rm A}=0.2$ (upper panels), $0.5$ (bottom panels) at viewing angle $\phi = 0^{\circ}$ (left panels), $60^{\circ}$ (right panels). The source is located at 2\,kpc away from Earth.}
\label{fig:map}
\end{figure*}

On the other hand, CR will contaminate the gamma-ray detection on the ground. At energy above 10\,TeV, the measurement of the muon component by LHAASO-KM2A provides an efficient rejection to the CR background and greatly improves the source detection capability. The simulated fraction of cosmic rays that pass the hadron rejection cuts can reach as low as $\sim 0.01\%$ at several tens TeV and even below $0.001\%$ above 100\,TeV \citep{2019arXiv190502773C}. The background counts rate above energy $E_\gamma$ per solid angle has been measured towards the direction of Crab Nebula \citep{2021Sci...373..425L}, as shown in Fig.~\ref{fig:performance}c. We denote the background counts rate by $B_{\rm CR}$ and assume it homogeneous over the entire sky given the highly isotropic distribution of the CR arrival direction measured locally.

\begin{figure*}[htbp]
\centering
\includegraphics[width=1\textwidth]{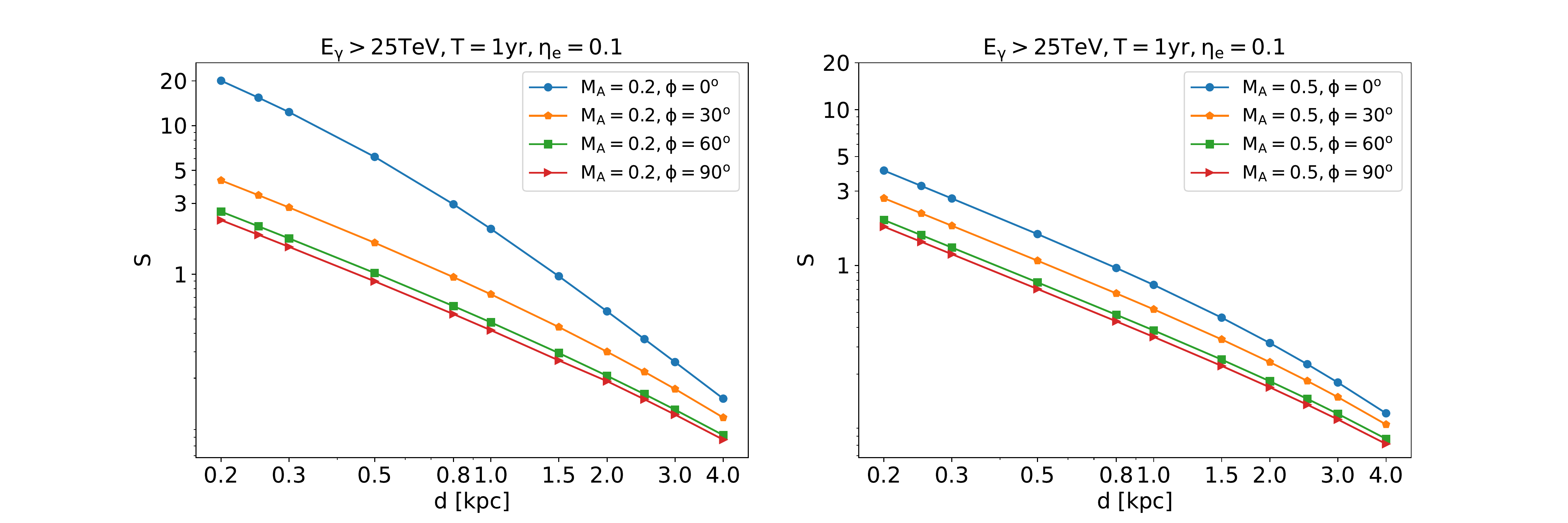}
\includegraphics[width=1\textwidth]{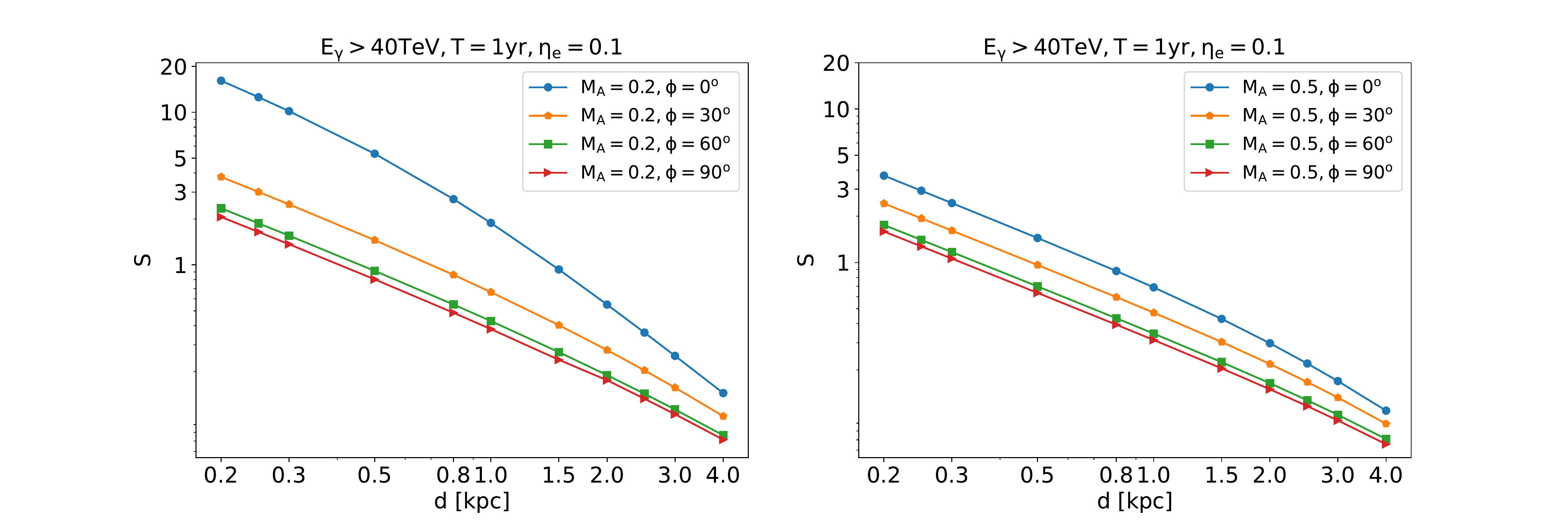}
\includegraphics[width=1\textwidth]{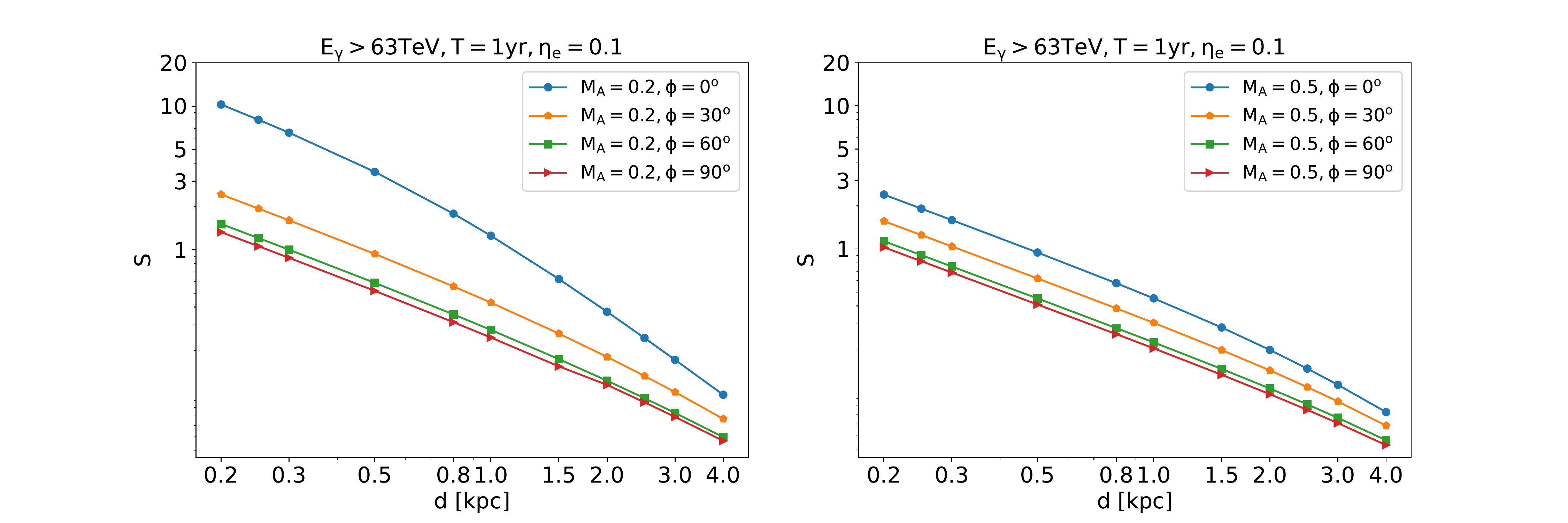}
\caption{Expected significance of Geminga-like halos at different distances with one-year operation of LHAASO. $M_{\rm A}=0.2$ in the left column, and $M_{\rm A}=0.5$ in the right column. Top, middle and bottom panels correspond to the cases with threshold energies of gamma-ray photons of 25\,TeV, 40\,TeV, 63\,TeV, respectively, employed in the analysis.}
\label{fig:snr}
\end{figure*}

We then estimate the SNR (or the statistical significance) of the pulsar halo with the Li-Ma formula \citep{1983ApJ...272..317L}, i.e., 
\begin{equation}\label{eq:snr}
\begin{split}
S&=\sqrt{2}\left\{
N_{\rm on} \ln \left[\frac{1+\alpha}{\alpha}\left(\frac{N_{\rm on}}{N_{\rm on}+N_{\rm off}}\right)\right] \right.\\
&\left. +N_{\rm off} \ln \left[(1+\alpha)\left(\frac{N_{\rm off}}{N_{\rm on}+N_{\rm off}}\right)\right]  \right\}^{1/2}
\end{split}
\end{equation}
where $\alpha$ is the ratio of the on-source time to the off-source time which is generally unity for LHAASO, and
\begin{eqnarray}
&& N_{\rm on}=(T/4)\int_0^{\theta_{\rm s}}\int_0^{2\pi} (0.7C_\gamma+B_{\rm CR})\sin\theta d\theta\xi\\
&& N_{\rm off}=(T/4)\int_0^{\theta_{\rm s}}\int_0^{2\pi} B_{\rm CR}\sin\theta d\theta\xi  
\end{eqnarray}
are the numbers of on-source and off-source counts respectively. The coefficient 0.7 before the source's counts rate $C_\gamma(>E_\gamma, \theta,\xi)$ accounts for the fact that about 30\% gamma-ray events can not survive after applying for the background rejection cuts. $T$ is the operation time period of LHAASO and we divide it by a factor of 4 assuming that each source appears 6 hours each day in LHAASO's field of view. By integrating the azimuthal $\xi$ over 0 to $2\pi$ in the above two equations,  we simply consider a symmetric disk with a radius of $\theta_s$ for the spatial template of the source, as commonly employed in practical data analyses of LHAASO and HAWC. This is sufficient for the goal of the present work, although a different spatial template could influence the resultant significance of the source as will be briefly discussed later. Note that a pulsar halo does not have a sharp boundary so we will search for the ``best-fit'' source size $\theta_{\rm s}$ by maximizing the significance of the source, as also did in practical data analyses. 

\section{Results} 

\begin{figure*}[htbp]
\centering
\includegraphics[width=1\textwidth]{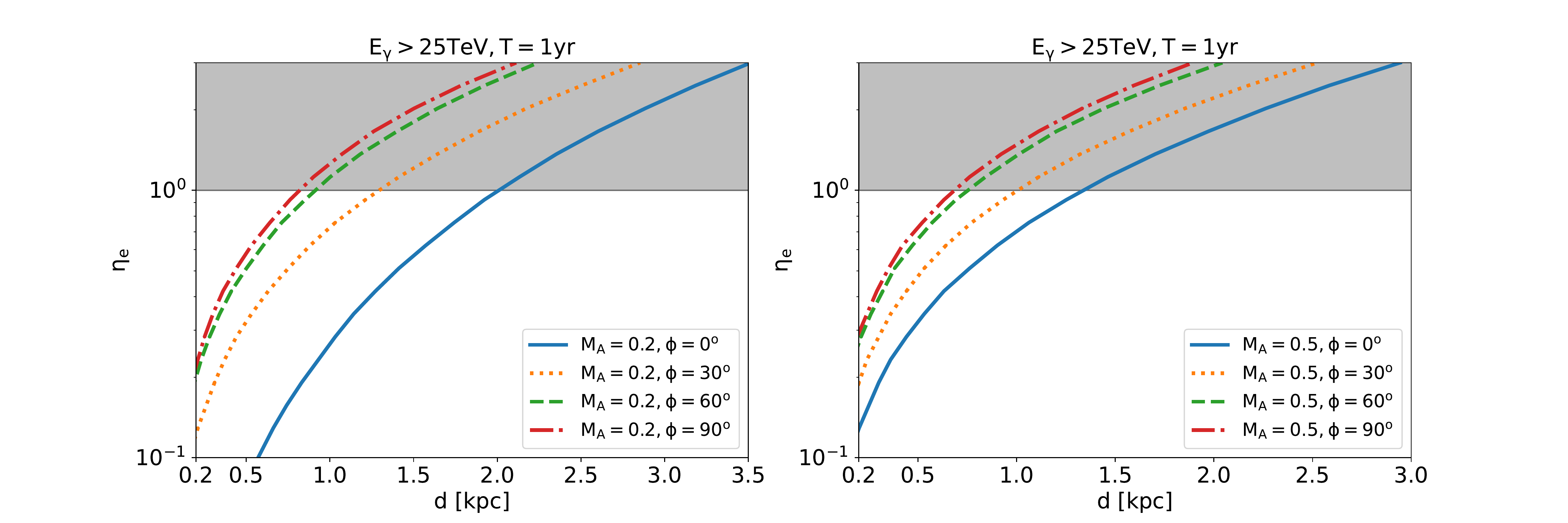}
\includegraphics[width=1\textwidth]{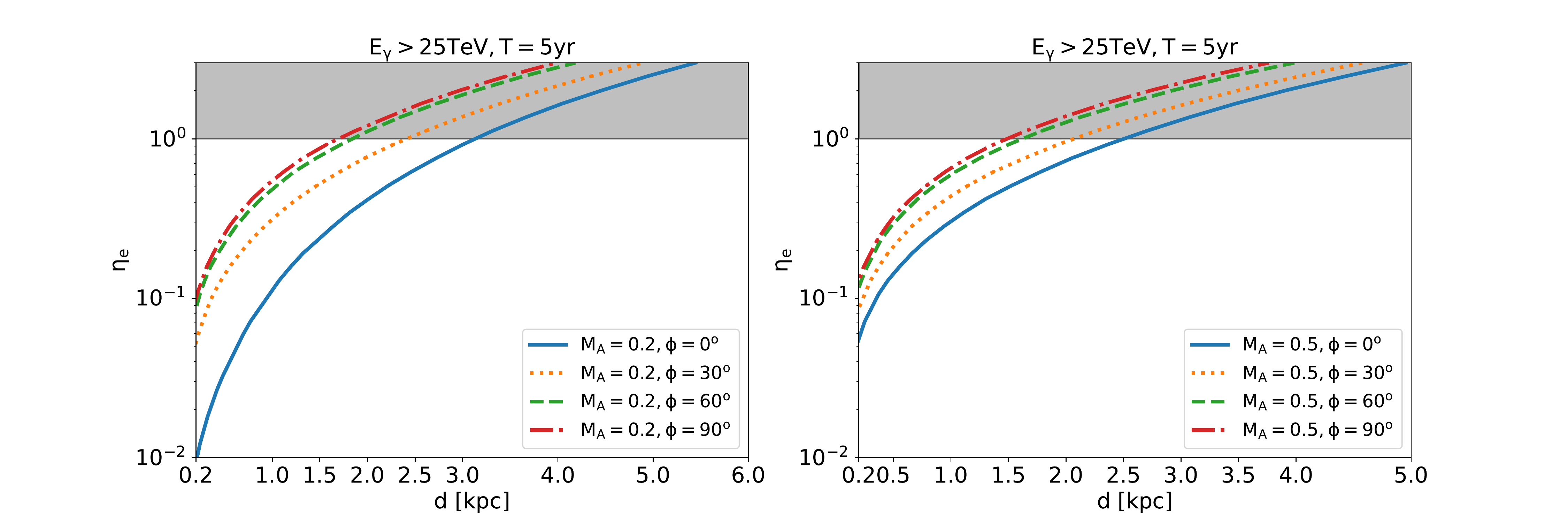}
\caption{Solid curves show the combination of the electron conversion efficiency $\eta_e$ and the source distance $d$ that can result in $5\sigma$ detection of hypothetical Geminga-like pulsar halos by LHAASO. Pulsar halos located on the left side of each curve can be detected  more significantly, while those on the right side of each curve cannot be detected. The grey shaded areas represent cases with unrealistic parameter $\eta_{\rm {e}} > 1$ which can be regarded as the requirement for a more powerful pulsar than Geminga. Operation time of LHAASO is set to be one year for upper panels and five years for the lower panels.}
\label{fig:boundary}
\end{figure*}

The main purpose of this work is to explore the unique feature of a pulsar halo expected in the anisotropic model, so that we can figure out a way to test the model or to distinguish between the anisotropic diffusion model and the isotropic diffusion model based on the future observation of LHAASO. We do not aim to simulate a pulsar halo identical to the observed one. Thus, we choose to fix some of the model parameters that are common in different models for simplicity, and focus on the influence of those unique parameters in the anisotropic model. In this spirit, we consider the pulsar halo produced by a Geminga-like pulsar, i.e., with the current spindown luminosity $L_s=3.3\times 10^{34}\rm erg/s$, characteristic age of 342\,kyr and the initial spindown timescale $\tau=12\,$kyr. The injection spectral index $p$ and the spectral cutoff energy are fixed at 2 and 200\,TeV respectively.

We show the spectral energy distribution (SED) of the simulated pulsar halo in Fig.~\ref{fig:spectrum} with assuming a typical conversion efficiency $\eta_e=0.1$. The spectrum does not depend on the viewing angle $\phi$ or the Alfv{\'e}nic Mach number $M_A$ after being integrated over the solid angle. On the other hand, the latter two parameters determine the spatial distribution of escaped electrons and consequently the morphology of the halo, as reflected in the 2D intensity map, i.e., $I_{\rm PSF}(E_\gamma, \theta,\xi)$, shown in Fig.~\ref{fig:map}.

%We quantitatively calculate LHAASO's detection abilities to Geminga-like TeV halos under the framework of anisotropic diffusion model. The variations of parameters are as follows: 1. the working time of LHAASO is set to be 1 year and 5 years, and the effective detection time is one quarter of the working time; 2. the integral energy ranges are set to be $E_{\gamma}> 25$ TeV, $E_{\gamma}> 40$ TeV and $E_{\gamma}> 63$ TeV; 3. the  $\mathrm{Alfv\acute{e}nic }$ Mach number $M_{\rm A}$ is set to be 0.2 and 0.5; 4. the distance from the simulated sources to the observer ranges from 0.3 kpc to 5 kpc; 5. the viewing angle between the Galactic mean magnetic field to LOS ranges from $0^\circ$ to $90^\circ$; 6. the energy conversion efficiency $\eta_{e}$ ranges from 0 to 1.

We then use a disk as the spatial template to search for the best-fit size of the halo and obtain the significance (or SNR) of the halo measured by LHAASO. In Fig.~\ref{fig:snr}, we present the significance of the halo as a function of distance from Earth, for $M_A=0.2$ (left panels) and $M_A=0.5$ (right panels) and for different viewing angles with different colors. The operation time of the instrument is taken to be one year. The total number of photon counts from the source depends on the threshold  energy that we start to involve. Although the number of photons is greater for a lower threshold energy, the CR background rate also increases by a lot. Also, the effective area of LHAASO-KM2A decreases significantly below 20\,TeV. So a smaller energy threshold does not necessarily result in a higher source significance, but depends on the source spectrum. We compare the result with $E_{\gamma}> 25$ TeV, $E_{\gamma}> 40$ TeV and $E_{\gamma}> 63$ TeV. For the employed injection spectrum with $p=2$, we see that simulated pulsar halos have the highest SNR with $E_{\rm \gamma}>25\,$TeV. Thus, we will focus on the result with $E_{\rm \gamma}>25\,$TeV in the following discussion.

The obtained statistical significance has a strong dependency on the source distance $d$, $M_{\rm A}$ value and viewing angle $\phi$. As expected, a smaller $M_{\rm A}$ leads to a slower perpendicular diffusion and hence the electron's spatial distribution is more compact. For a smaller viewing angle $\phi$, the projection of the halo on the celestial sphere appears more isotropic and the apparent morphology is more concentrated. Therefore, either a smaller $M_{\rm A}$ or a smaller $\phi$ would lead to a more compact source with the same total luminosity. In other words, while $N_{\rm on}$ may keep more or less the same, $N_{\rm off}$ can be reduced given a smaller $M_{\rm A}$ and $\phi$, resulting in an enhancement of the significance of detection. By contrast, a larger $M_{\rm A}$ and $\phi$ would make the source emission more diluted and dimmer. Note that the mean magnetic field direction (related to $\phi$) and the ISM turbulent level (related to $M_{\rm A}$) may vary from place to place. Consequently, a pulsar with a higher spindown power and/or located at a smaller distance does not guarantee a more significant detection.

In general, we may claim discovery of a source if the significance exceeds $5\sigma$ confidence level (or the SNR is greater than 5). From the figure we can see that for $M_A=0.2$, the hypothetical pulsar halo cannot be discovered with one-year operation of LHAASO if the viewing angle $\phi$ is larger than $30^\circ$. The detection becomes more difficult if the ISM around the pulsar is more turbulent with $M_A=0.5$. The pulsar halo cannot be significantly detected with one-year operation of LHAASO even for $\phi=0^\circ$, unless it is located closer than 200\,pc from Earth, or the electron injection rate is higher (i.e., a larger electron conversion efficiency $\eta_e$). 

The influence of the source distance and the electron conversion efficiency on the detection of the pulsar halo can be further investigated. Given the turbulent level (i.e., $M_{\rm A}$) and the mean field direction ($\phi$) of the interstellar magnetic field around a pulsar, we can find out the critical combination of the distance and the electron conversion efficiency for a $5\sigma$ detection significance of the pulsar halo. The result is shown in Fig.~\ref{fig:boundary} with different curves representing different $M_A$ and $\phi$. Pulsar halos located on the left side of each curve in the $\eta_e-d$ parameter space can lead to detection of the halo by LHAASO-KM2A with one-year operation, while those on the right side of the curve cannot be detected. For a typical electron conversion efficiency $\eta_e=0.1$, a Geminga-like halo can be detected only if it is located within about 0.6\,kpc even for $M_A=0.2$ and $\phi=0^\circ$, which is the most optimistic condition considered in our simulation. For the most extreme case with $\eta_e=1$, the halo can be detected up to $d\lesssim 2\,$kpc. On the other hand, it is difficult to detect those pulsar halos with larger viewing angles, i.e., $\phi>30^\circ$, at $d>1\,$kpc, unless they are more powerful than the Geminga pulsar as represented with $\eta_e>1$ (see the shaded region in Fig.~\ref{fig:boundary}). 

A longer exposure time can increase the SNR, roughly scaling  with $\sqrt{T}$. We show the result with five-year operation of LHAASO in the lower panels of Fig.~\ref{fig:boundary}. It indeed reduce the required electron power or $\eta_e$ or increase the detectable distance for the $5\sigma$ detection of the pulsar halo. This is helpful since it can increase the number of potential sources and enhance the possibility to discover an elongated pulsar halo.

\begin{figure*}[htbp]
\centering
\includegraphics[width=0.48\textwidth]{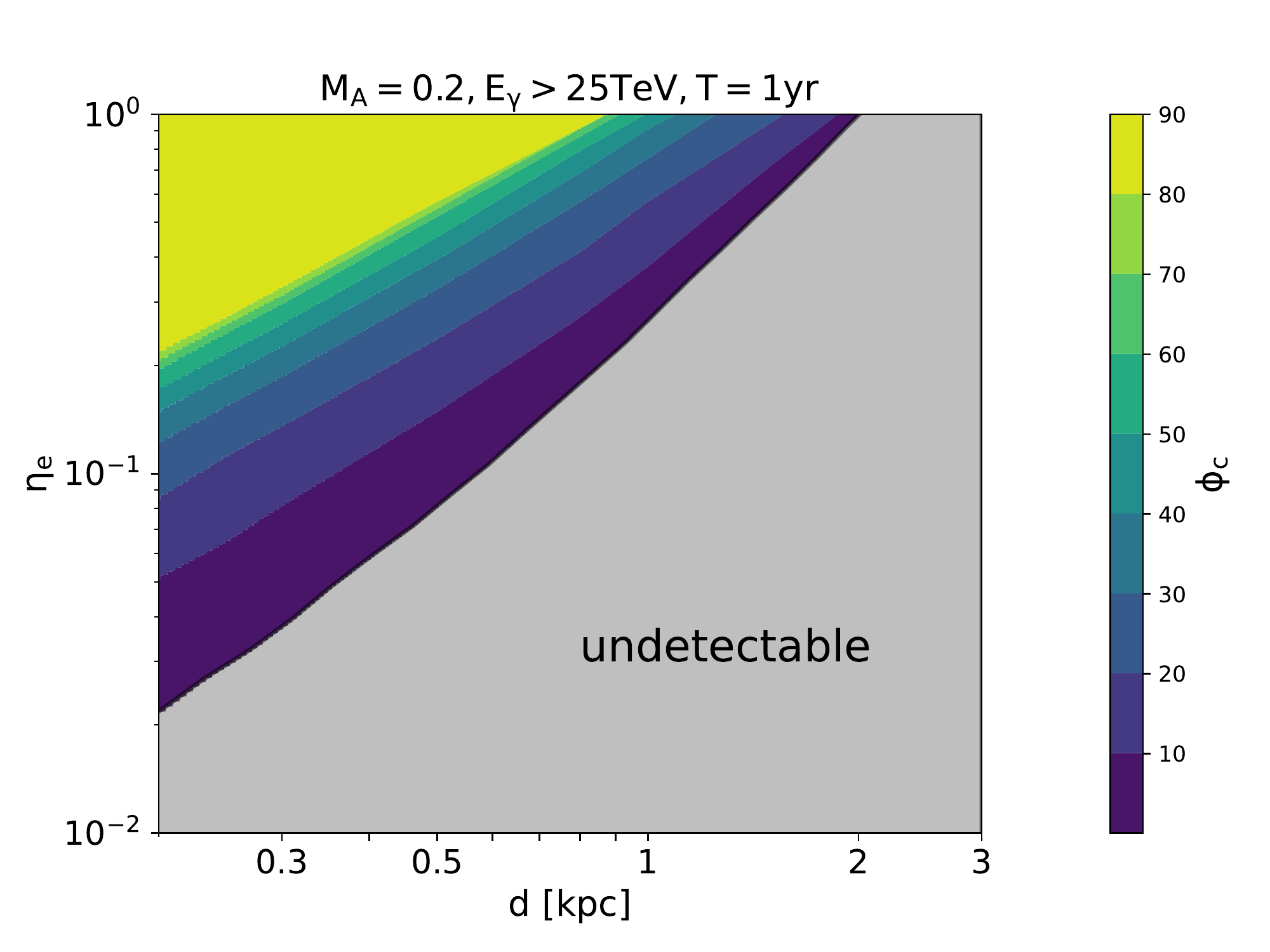}
\includegraphics[width=0.48\textwidth]{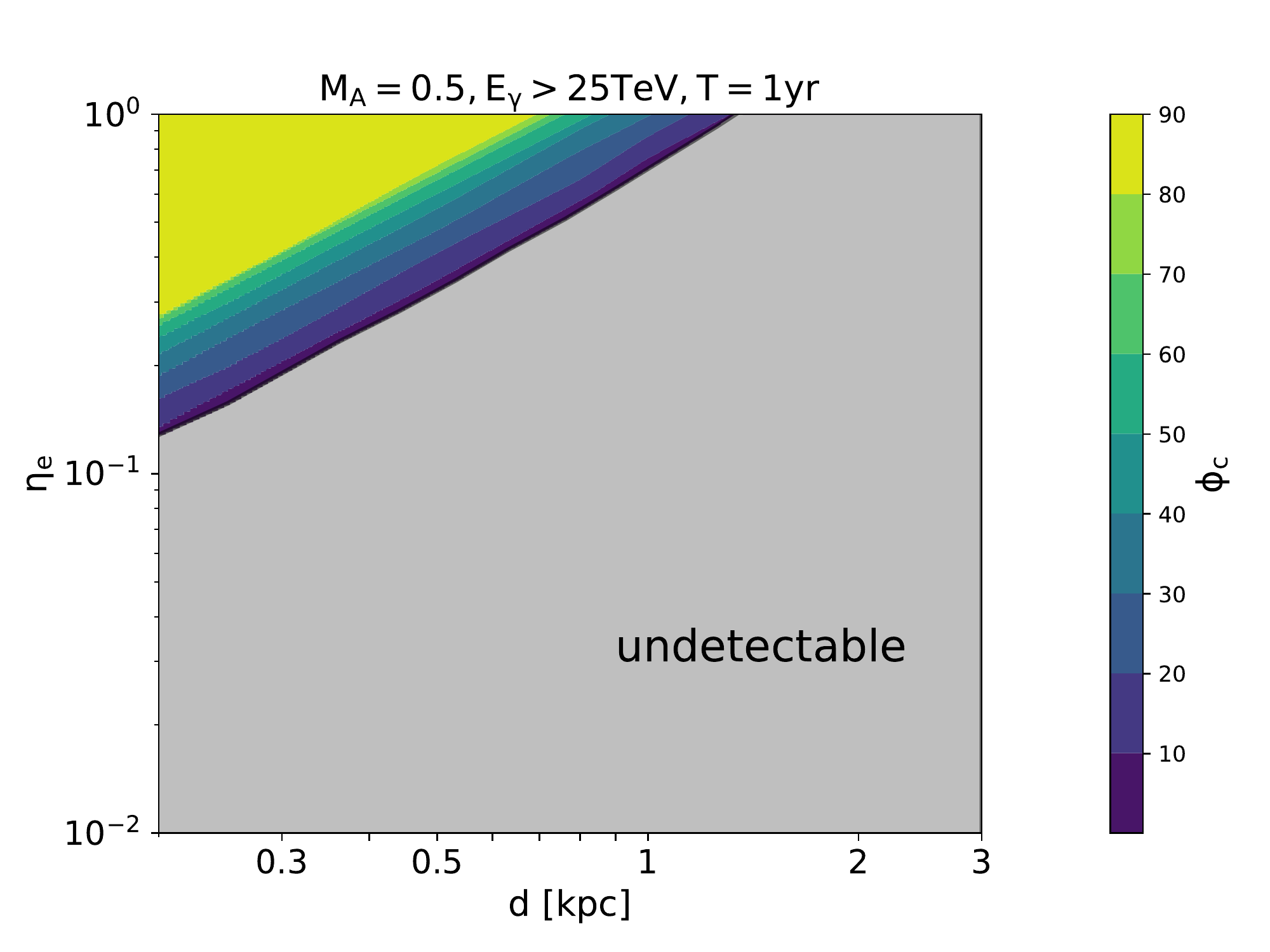}
\includegraphics[width=0.48\textwidth]{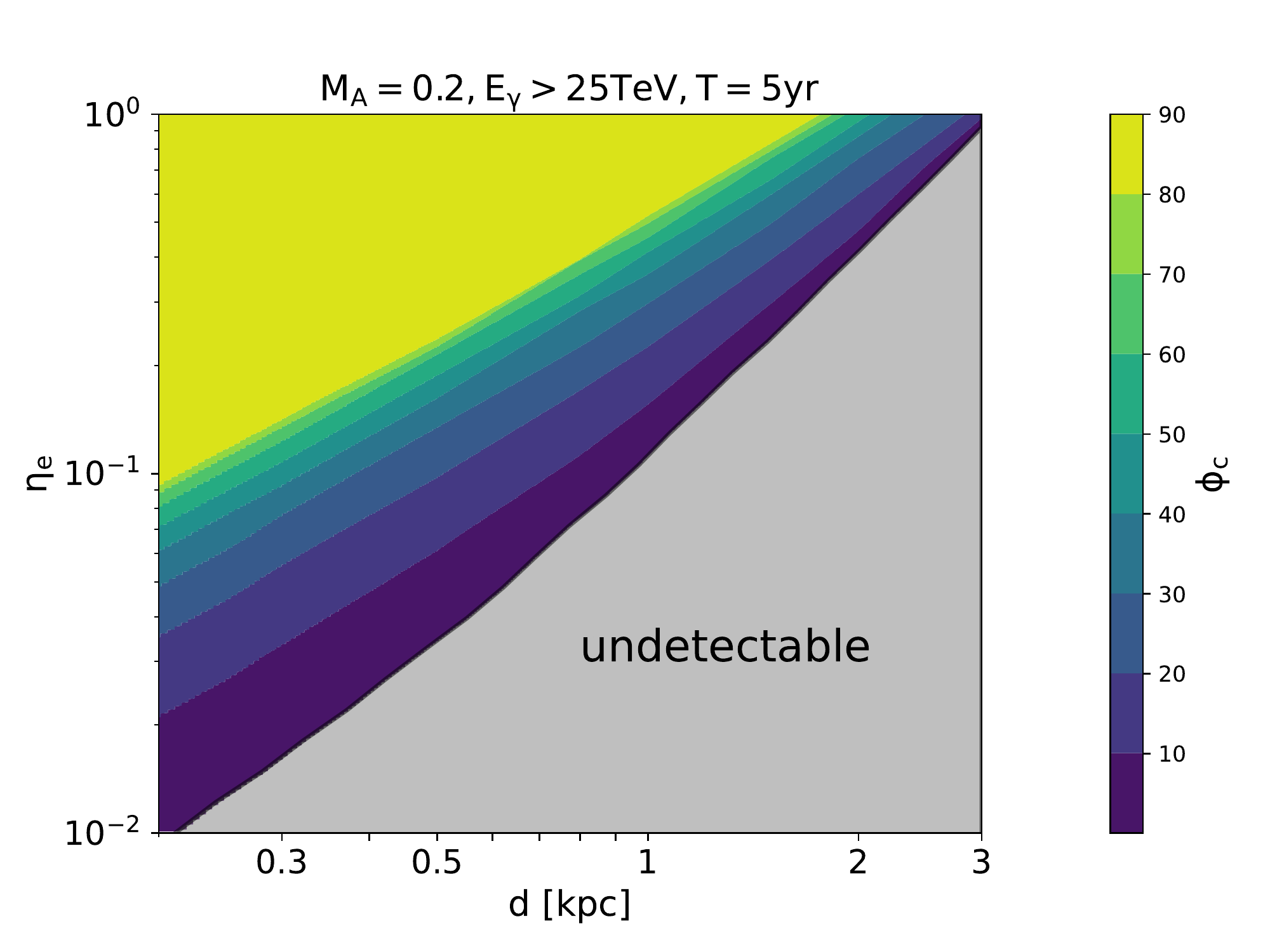}
\includegraphics[width=0.48\textwidth]{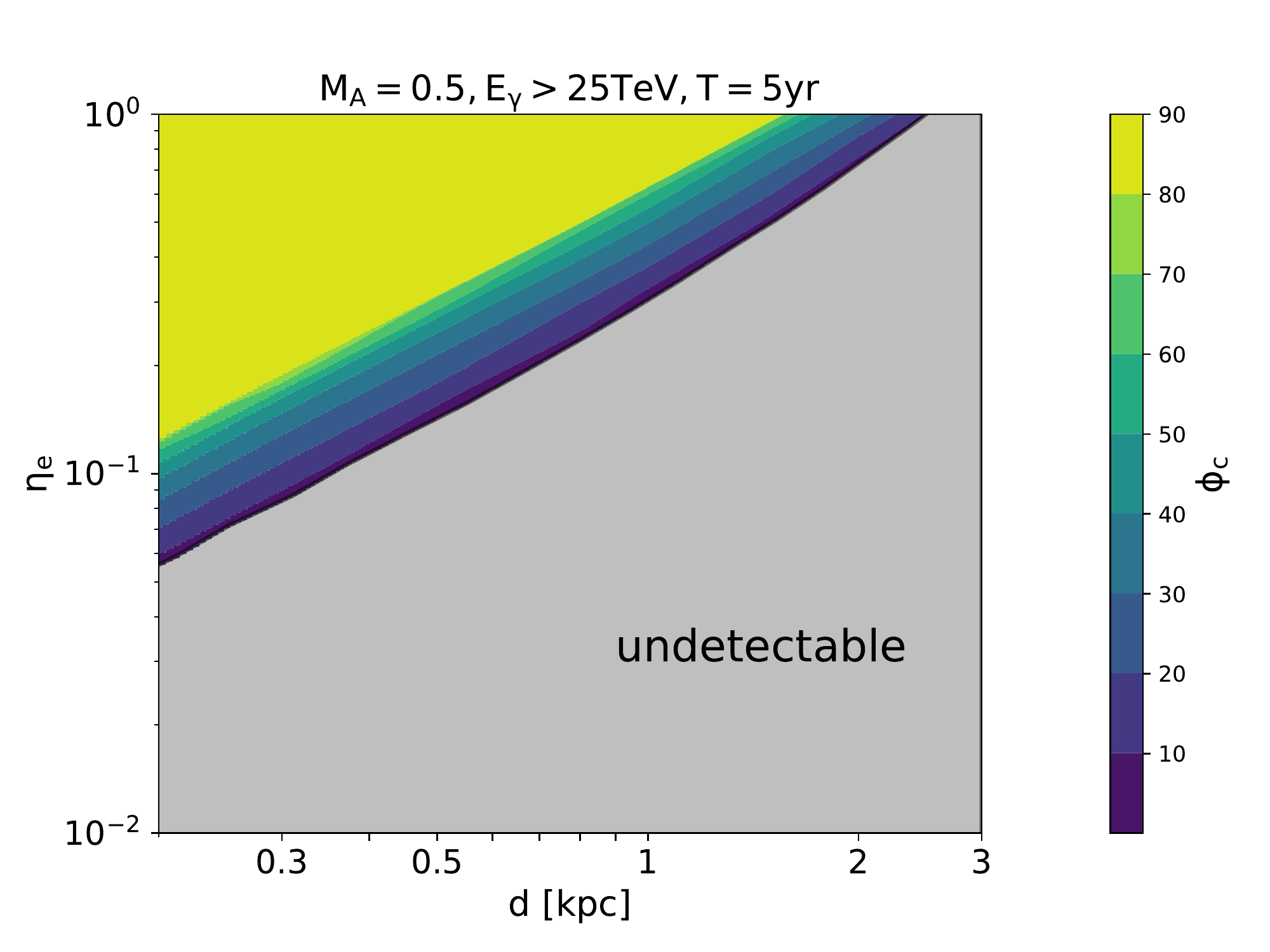}
\caption{Distribution of the critical viewing angle $\phi_c$ in the $\eta_e-d$ space. The value of $\phi_c$ is represented by different colors. The hypothetical pulsar halo with a viewing angle smaller than $\phi_c$ can be discovered by LHAASO with a statistical significance greater than $5\sigma$. The grey regions represent the combination of $\eta_e$ and $d$ with which the halo cannot be detected. The boundary separating detectable region (the dark blue region) and the undetectable region (the grey region) is identical to the blue curves shown in Fig.~\ref{fig:boundary}.}
\label{fig:angle_dis}
\end{figure*}

To quantify the influence of the viewing angle on the detection of pulsar halos, we define the critical viewing angle $\phi_c$, within which $\geq 5\sigma$ significance can be achieved for a specific combination of $\eta_e$, $d$ and $M_A$. 
Fig.~\ref{fig:angle_dis} shows the distribution of the critical viewing angle $\phi_c$ in the $\eta_{e}-d$ plane for $M_A=0.2$ (left panels) and $M_A=0.5$ (right panels), as well as for an operation time of $T=1\,$yr (upper panels) and $T=5\,$yr (lower panels). 
We see that in the case of $M_A=0.2$ and $T=5\,$yr, it is possible to detect those highly elongated halos up to $d\approx 1\,$kpc with a reasonable conversion efficiency of $\eta_e\approx 0.5$, and 
those with moderate asymmetric morphology, i.e., $\phi=10^\circ-20^\circ$, up to $d\lesssim 2\,$kpc with a more relaxing conversion efficiency of $\eta_e\gtrsim 0.1$. However, we caveat that a highly elongated pulsar halos could overlap with other sources and hence their identification may be a challenge in the practical data analysis.  

Comparing the cases of $T=1\,$yr and $T=5\,$yr, we can envisage that the detectable region in the $\eta_e-d$ space gradually expands downwards to the right in the panel with the operation time, and the value of the critical viewing angle $\phi_c$ for certain combination of $\eta_e$ and $d$ also gradually increases with time. Therefore, in the anisotropic diffusion model, we expect to firstly detect pulsar halos with small viewing angles, if there is any, and then detect those with larger viewing angles. Such a selection effect may explain why the three identified pulsar halos as of now do not show significant anisotropic morphology.

\section{Discussion}
The main goal of this paper is to discuss the influence of the viewing angle $\phi_c$ on the detectability of pulsar halos, and hence we fix many model parameters for simplicity. However, some of these parameters can also influence the predicted SNR of the pulsar halo and we briefly discuss them in this section.

\subsection{Injection spectrum of electrons}
In our simulation shown in the last section, the spectral index of the injection electrons $p$ is assumed to be 2 and the cutoff energy $E_{\rm max}$ is assumed to be 200\,TeV. The injection electron spectrum may differ for different sources and are not well constrained even for the observed halos. A different injection spectrum can affect the gamma-ray flux of the hypothetical pulsar halo and consequently the significance of the source. From Fig.~\ref{fig:spectrum}, we see that a harder/softer injection electron spectrum results in a higher/lower flux above 25\,TeV. For instance, if we take $p=1.6$ as the slope of the injection spectrum, the number of the signal counts can be increased by a factor of 2 with respect to currently employed $p=2$ case. Using a larger/smaller $E_{\rm max}$ can also increase/decrease the flux above 25\,TeV, but its influence is within a factor of 2 as long as $E_{\rm max}>100\,$TeV.

\subsection{Pulsar's properties}
In this work, we consider the pulsar's properties such as the period, the age and the spindown power to be the same as the Geminga pulsar. The initial rotation period of the pulsar and the braking index are fixed to be 50\,ms and 3 respectively, as they cannot be determined by observation. Although these parameters are different among different pulsars, their influence on the result is not important. This is because we focus on the detection by LHAASO-KM2A which mainly works at the energy above a few tens of TeV. At such high energies, photons mainly originate from the radiation of $\gtrsim 100$\,TeV electrons via up-scattering CMB. The cooling timescale of such a high-energy electron is only $t_{\rm cool}\approx
5(E_e/100{\rm TeV})^{-1}(B/5\mu \rm G)^{-2}\,$kyr, implying the emitting electrons are injected very recently with respect to the pulsar's age of $\sim 100\,$kyr. As a result, the spindown history of the pulsar does not affect the halo's emission of the present time at such high energies.

On the other hand, the resulting gamma-ray flux depends on the pulsar's current spindown power linearly. This influence is straightforward and can be referred to that of $\eta_{e}$ as shown in Section 3.

% The Galactic mean magnetic field can dominate the diffusion of a TeV halo only when the PWN has evolved far enough away from the supernova remnant or the supernova remnant has alreadly disappeared, which is also the premise of the anisotropic diffusion model. This means that the anisotropic diffusion model does not apply to young PWNe, which do not exhibit TeV halo activity or whose diffusion and radiation are isotropic \citep{PhysRevD.100.043016}. In this work, we limit to discuss the Geminga-like sources and do not involve the energy transportation progress in young PWNe. 

% In this work, the spin-down timescale of pulsars are set to be 12 kyr for the purpose to study Geminga-like sources. This value should be adjusted when analysing other specific sources, which also causes a change in the detection rate.

\subsection{Spatial template used for analysis}
As is shown, it is more difficult to detect pulsar halos with larger viewing angles. On the one hand, the larger viewing angle makes the halo more elongated in the sky, diluting the intensity. On the other hand, the signal of the source with the asymmetric morphology cannot be well extracted with a symmetric disk, which is employed as the spatial template in our analysis. The latter may be improved by using an elliptical spatial template for the analysis in principle. Intuitively, an elliptical template adapted to the morphology of the elongated halo can better focus on the source region and involve less background. On the other hand, the elliptical template has two more degrees of freedom, i.e., the ratio of the long axis to the short axis and the inclination angle, compared to the symmetric templates such as a symmetric 2D Gaussian function or a spherical disk. In practical data analysis, the obtained statistical significance of the source will suffer a penalty if the template has more degrees of freedom. Therefore, using an elliptical template may not necessarily enhance the significance if the statistics is not sufficiently high. A detailed discussion on the spatial template is, however, beyond the scope of this paper, and we leave the influence of spatial templates for the future study. 

%For ideal sources of numerical simulation, the ellipticity of the optimal elliptical template is roughly the ellipticity of the morphology of the TeV halo projected perpendicular to the LOS. After convolving with the PSF of the instrument, the ellipticity of the optimal elliptical template will reduce to some extent. From another point of view, this ellipticity can be used to estimate the direction of the Galactic mean magnetic field. Since the current mainstream approach is to use disks, the application of elliptical templates still needs to be considered in combination with specific data.

\section{Summary}
The formation of TeV pulsar halos is not fully understood yet. The observed features may be reproduced either with the isotropic diffusion model or the anisotropic diffusion model. In this work, we studied LHAASO's detection ability for TeV pulsar halos under the framework of the anisotropic diffusion model. We firstly simulated the 2D intensity profile of a Geminga-like pulsar halo under different key parameters such as the viewing angle (defined as the angle between the observer's line of sight to the pulsar and the mean magnetic field direction of the surrounding ISM) and the Alfv{\'e}nic Mach number. We converted the intensity map to the counts map as if it were measured by LHAASO-KM2A based on the latter's effective area, PSF and the noise rate. We then used the Li-Ma formula to calculate the signal-to-noise ratio or the statistical significance of the pulsar halo detected by LHAASO. We showed that the viewing angle of the pulsar halo and the Alfv{\'e}nic Mach number have important influences on the morphology of the halo as well as the expected signal-to-noise ratio. More specifically, a smaller $M_{\rm A}$ and/or a smaller $\phi$ result in a more compact source and lead to a higher signal-to-noise ratio of the source, and vice versa. Given one-year operation of LHAASO, it is difficult to observe a highly elongated pulsar halo (i.e., with a large viewing angle $\phi$), unless it is several times more powerful as a relativistic electron emitter than Geminga or it is located very close to us such as within 200\,pc. Such a selection effect may explain why none of the three detected pulsar halos up to date shows strong asymmetric morphology. Our simulations also implied that LHAASO can hopefully detect Geminga-like halo with a large viewing angle after a reasonably long exposure time, such as five years, if the electron conversion efficiency is not too low (i.e., $>0.1$).
The halo with a large viewing angle would appear asymmetric in the
sky, which is not expected by the isotropic diffusion model. On the other hand, it'd rule out the anisotropic diffusion model if not a single pulsar halo with the asymmetric morphology is detected by LHAASO in a several years. Hence, the search for those asymmetric or elongated pulsar halos can serve as a critical test for the anisotropic diffusion model and provide a clue to understand the cosmic-ray transport mechanism in the ISM.

% conclude that assuming the energy conversion efficiency $\eta_{e}$ is $50\%$, working for 5 years, LHAASO will be able to detect all Geminga-like TeV halos within 1 kpc with any Galactic mean magnetic field orientation for both $M_{\rm A}=0.2$ and 0.5. This indicates that LHAASO can provide good tests for the anisotropic diffusion model in the future.

% Also, we note that sources those with LOS more aligned to the ambient mean magnetic field are easier to detect, thus resulting in the instrumental selection effects. We analyze influences of various parameters on the detection ability to anisotropic TeV halos, and found that the energy range $E_{\gamma}>25$ TeV is the most favorable and the energy conversion efficiency $\eta_{e}$ has a strong influence. Besides, model uncertainties due to the differences 
% in the physical states of pulsars might also affect the detection rate. Although more difficult to detect, TeV halos of anisotropic morphology should be observed after a long exposure time. LHAASO observation would substantially confirm or rule out the anisotropic diffusion model and help to reveal the transport mechanism in TeV halos. 

\section*{Acknowledgements}
We thank Hai-Ming Zhang and Yi Zhang for the helpful discussions. This work is supported by the NSFC grant No.U2031105 and No.12022502.

\bibliography{ms}
\bibliographystyle{aasjournal}

\end{document}